# Asymptotic Analysis of Multicell Massive MIMO over Rician Fading Channels

Luca Sanguinetti*†, Abla Kammoun‡, Merouane Debbah†§

*Dipartimento di Ingegneria dell'Informazione, University of Pisa, Pisa, Italy
†Large Networks and System Group (LANEAS), CentraleSupélec, Université Paris-Saclay, Gif-sur-Yvette, France
‡ Electrical Engineering Department, King Abdullah University of Science and Technology, Thuwal, Saudi Arabia
§Mathematical and Algorithmic Sciences Lab, Huawei France, Paris, France

***Abstract*—This work considers the downlink of a multicell massive MIMO system in which $L$ base stations (BSs) of $N$ antennas each communicate with $K$ single-antenna user equipments randomly positioned in the coverage area. Within this setting, we are interested in evaluating the sum rate of the system when MRT and RZF are employed under the assumption that each intracell link forms a MIMO Rician fading channel. The analysis is conducted assuming that $N$ and $K$ grow large with a non-trivial ratio $N/K$ under the assumption that the data transmission in each cell is affected by channel estimation errors, pilot contamination, and an arbitrary large scale attenuation. Numerical results are used to validate the asymptotic analysis in the finite system regime and to evaluate the network performance under different settings. The asymptotic results are also instrumental to get insights into the interplay among system parameters.**

## I. Introduction

Massive MIMO (also known as large scale MIMO) is considered as one of the most promising technology for next generation cellular networks [1]–[4]. The massive MIMO technology aims at evolving the conventional base stations (BSs) by using arrays with a hundred or more small dipole antennas. This allows for coherent multi-user MIMO transmission where tens of users can be multiplexed in both the uplink (UL) and downlink (DL) of each cell. It is worth observing that, contrary to what the name "massive" suggests, massive MIMO arrays are rather compact; 160 dual-polarized antennas at 3.7 GHz fit into the form factor of a flat-screen television [5].

In this work, we consider the DL of a massive MIMO system in which $L$ BSs of $N$ antennas each communicate with $K$ single-antenna user equipments (UE) randomly positioned in the coverage area. We assume that the system is affected by channel estimation errors, pilot contamination, and an arbitrary large scale attenuation. Maximum ratio transmit (MRT) or regularized zero forcing (RZF) are used as linear precoding techniques. Differently from most of the existing literature, we model the intracell communication links as MIMO Rician fading, which is more general and accurate to capture the fading variations when there is a line-of-sight (LOS) component. Compared to the Rayleigh fading channel, a Rician model makes the analysis of massive MIMO systems much more involved. To overcome this issue, recent results from random matrix theory and large system analysis [6], [7] are used to compute asymptotic expressions of the signal-to-interference-plus-noise ratios (SINRs), which are eventually used to approximate the ergodic sum rates of the system. As a notable outcome of this work, the above analysis provides an analytical framework that can be used to evaluate the performance of the network under different settings without resorting to heavy Monte Carlo simulations and to eventually get insights on how the LOS components affect the network performance.

The main literature related to this work is represented by [3], [8]–[11]. Tools from random matrix theory are used in [8] to compute the ergodic sum rate in a single-cell MIMO setting with Rayleigh fading and different precoding schemes while the multicell case is analyzed in [9]. Similar tools are used in [10] to solve the power minimization problem under different configurations of cooperation among BSs. A similar large system analysis is presented in [3] for the UL and DL of Massive MIMO in cellular networks, wherein channel estimation and pilot contamination are also taken into account. All these works relies on random matrix theory but assume that a Rayleigh fading channel model. In [12], the authors investigate a LOS-based conjugate beamforming transmission scheme and derive some expressions of the statistical SINR under the assumption that $N$ grows large and $K$ is fixed. In [13], the authors study the fluctuations of the mutual information of a cooperative small cell network operating over a Rician fading channel under the form of a central limit theorem and provide an explicit expression of the asymptotic variance. In [14], a deterministic equivalent of the ergodic sum rate and an algorithm for evaluating the capacity achieving input covariance matrices for the uplink of a large-scale MIMO are proposed for spatially correlated MIMO channel with LOS components. The analysis of the uplink rate with both zero-forcing and maximum ratio combining receivers is performed in [15].

The remainder of this paper is organized as follows.[1] Next

This research has been supported by the ERC Starting Grant 305123 MORE, and by the research project 5GIOTTO funded by the University of Pisa.

[1]The following notation is used throughout the paper. Scalars are denoted by lower case letters whereas boldface lower (upper) case letters are used for vectors (matrices). We denote by $\mathbf{I}_N$ the identity matrix of order $N$ and call $[\mathbf{A}]_{ik}$ the $(i,k)$th element of the enclosed matrix. A random vector $\mathbf{x} \sim \mathcal{CN}(\mathbf{m}, \mathbf{C})$ is complex Gaussian distributed with mean $\mathbf{m}$ and covariance matrix $\mathbf{C}$. The trace, transpose, conjugate transpose, and expectation operators are denoted by $\mathrm{tr}$, $^T$, $^H$ and $\mathbb{E}[\cdot]$.

section introduces the system model for DL. The asymptotic analysis is performed in Section III and validated by means of numerical results in Section IV. Conclusions are drawn in Section V.

## II. System Model

Consider a multi-cell multi-user MIMO system composed of $L$ cells, the BS of each cell comprising $N$ antennas to communicate with $K$ single-antenna UEs. A double index notation is used to refer to each UE as e.g., "user $k$ in cell $j$". Under this convention, let $\mathbf{h}_{jlk} \in \mathbb{C}^N$ be the channel from BS $j$ to UE $k$ in cell $l$ within a block and assume that

$$\mathbf{h}_{jlk} = \sqrt{\beta_{jlk}}\mathbf{w}_{jlk} \tag{1}$$

where $\beta_{jlk}$ accounts for the corresponding large scale channel fading or path loss (from BS $j$ to UE $k$ in cell $l$) and $\mathbf{w}_{jlk} \in \mathbb{C}^N$ is the small scale fading channel. The channel matrix $\mathbf{H}_{jl} \in \mathbb{C}^{N\times K}$ from BS $l$ to BS $j$ is thus given by $\mathbf{H}_{jl} = [\mathbf{h}_{jl1}, \ldots, \mathbf{h}_{jlK}]$. We assume that

$$\mathbf{w}_{jjk} = \sqrt{\frac{1}{1+\kappa_{jk}}}\mathbf{z}_{jjk} + \sqrt{\frac{\kappa_{jk}}{1+\kappa_{jk}}}\mathbf{a}_{jjk} \qquad l = j \tag{2}$$

$$\mathbf{w}_{jlk} = \mathbf{z}_{jlk} \qquad l \neq j \tag{3}$$

where $\mathbf{z}_{jlk} \in \mathbb{C}^N$ is assumed to be Gaussian with zero mean and unit covariance, i.e., $\mathbf{z}_{jlk} \sim \mathcal{CN}(\mathbf{0}_N, \mathbf{I}_N)$, $\mathbf{a}_{jjk} \in \mathbb{C}^N$ is a deterministic vector, and the scalar $\kappa_{jk} \geq 0$ is the Rician factor denoting the ratio between $\mathbf{a}_{jjk}$ and $\mathbf{z}_{jjk}$. For notational convenience, we let

$$d_{jjk} = \frac{\beta_{jjk}}{1+\kappa_{jk}} \qquad l = j \tag{4}$$

$$d_{jlk} = \beta_{jlk} \qquad l \neq j \tag{5}$$

such that $\mathbf{h}_{jjk}$ can be rewritten as

$$\mathbf{h}_{jjk} = \widetilde{\mathbf{h}}_{jjk} + \overline{\mathbf{h}}_{jjk} \tag{6}$$

$$\mathbf{h}_{jlk} = \widetilde{\mathbf{h}}_{jlk} \tag{7}$$

with $\widetilde{\mathbf{h}}_{jlk} = \sqrt{d_{jlk}}\mathbf{z}_{jlk}$ and $\overline{\mathbf{h}}_{jjk} = \sqrt{d_{jjk}\kappa_{jk}}\mathbf{a}_{jjk}$.

### *A. Channel estimation*

We assume that the BS and UEs are perfectly synchronized and operate according to a time-division duplex (TDD) protocol wherein the DL data transmission phase is preceded in the UL by a training phase for channel estimation. If a single-cell MMSE estimator is employed [3], then the estimate $\widehat{\mathbf{h}}_{jji}$ of $\mathbf{h}_{jjk}$ $\forall j,k$ is given by

$$\widehat{\mathbf{h}}_{jjk} = \overline{\mathbf{h}}_{jjk} + \frac{d_{jjk}}{\frac{1}{\rho^{\text{tr}}} + \sum\limits_{n=1}^{L} d_{jnk}}\left(\mathbf{y}_{jk}^{\text{tr}} - \overline{\mathbf{h}}_{jjk}\right) \tag{8}$$

where $\rho^{\text{tr}}$ accounts for the SNR during the UL training phase and $\mathbf{y}_{jk}^{\text{tr}}$ is given by

$$\mathbf{y}_{jk}^{\text{tr}} = \mathbf{h}_{jjk} + \sum_{l=1, l\neq j}^{L} \mathbf{h}_{jlk} + \frac{1}{\sqrt{\rho^{\text{tr}}}}\mathbf{n}_{jk}^{\text{tr}} \tag{9}$$

with $\mathbf{n}_{jk}^{\text{tr}} \sim \mathcal{CN}(\mathbf{0}_N, \mathbf{I}_N)$. The estimate $\widehat{\mathbf{h}}_{jjk}$ is distributed as $\widehat{\mathbf{h}}_{jjk} \sim \mathcal{CN}(\overline{\mathbf{h}}_{jjk}, \phi_{jjk}\mathbf{I}_N)$ with

$$\phi_{jlk} = \frac{d_{jjk}d_{jlk}}{\frac{1}{\rho^{\text{tr}}} + \sum\limits_{n=1}^{L} d_{jnk}}. \tag{10}$$

The estimated uplink channel of cell $j$ is thus given by $\widehat{\mathbf{H}}_{jj} = [\widehat{\mathbf{h}}_{jj1}, \ldots, \widehat{\mathbf{h}}_{jjK}]$. According to the orthogonality principle of MMSE estimation, the estimation error $\mathbf{e}_{jjk} = \mathbf{h}_{jjk} - \widehat{\mathbf{h}}_{jjk}$ is distributed as $\sim \mathcal{CN}\left(\mathbf{0}_N, (d_{jjk} - \phi_{jjk})\,\mathbf{I}_N\right)$.

### *B. Achievable rates with linear precoding*

We denote by $\mathbf{g}_{jk} \in \mathbb{C}^N$ the precoding vector of UE $k$ in cell $j$. As in [1], [3], [16], [17] (among many others), we assume that there are no downlink pilots such that the UEs do not have knowledge of the current channels but can only learn the average channel gain $\mathbb{E}\{\mathbf{h}_{jjk}^H\mathbf{g}_{jk}\}$ and the total interference power. Using the same technique from [18], an ergodic achievable information rate for UE $k$ in cell $j$ is obtained as

$$r_{jk} = \log_2(1+\gamma_{jk}) \tag{11}$$

where $\gamma_{jk}$ is given by

$$\gamma_{jk} = \frac{|\mathbb{E}[\mathbf{h}_{jjk}^H\mathbf{g}_{jk}]|^2}{\frac{1}{\rho^{\text{dl}}} + \sum\limits_{l=1}^{L}\sum\limits_{i=1}^{K}\mathbb{E}[|\mathbf{h}_{ljk}^H\mathbf{g}_{li}|^2] - |\mathbb{E}[\mathbf{h}_{jjk}^H\mathbf{g}_{jk}]|^2} \tag{12}$$

where the expectation is taken with respect to the channel realizations and $\rho^{\text{dl}}$ accounts for the SNR in the DL. The above result holds true for any precoding scheme and is obtained by treating the inter-user interference (from the same and other cells) and channel uncertainty as worst-case Gaussian noise. As mentioned earlier, we consider MRT and RZF as precoding schemes [1], [3], [4], [16]. This yields

$$\mathbf{g}_{jk}^{\text{MRT}} = \frac{\widehat{\mathbf{h}}_{jjk}}{\sqrt{\mathbb{E}\left[\frac{1}{K}\sum\limits_{k=1}^{K}||\widehat{\mathbf{h}}_{jjk}||^2\right]}} = \sqrt{\theta_j}\widehat{\mathbf{h}}_{jjk} \tag{13}$$

$$\mathbf{g}_{jk}^{\text{RZF}} = \frac{\widehat{\mathbf{u}}_{jk}}{\sqrt{\mathbb{E}\left[\frac{1}{K}\sum\limits_{k=1}^{K}||\widehat{\mathbf{u}}_{jk}||^2\right]}} = \sqrt{\psi_j}\widehat{\mathbf{u}}_{jk} \tag{14}$$

where $\widehat{\mathbf{u}}_{jk} = \frac{1}{N}\mathbf{Q}_j\widehat{\mathbf{h}}_{jjk}$ with

$$\mathbf{Q}_j = \left(\frac{1}{N}\sum_{i=1}^{K}\widehat{\mathbf{h}}_{jji}\widehat{\mathbf{h}}_{jji}^H + \lambda_j^{\text{dl}}\mathbf{I}_N\right)^{-1}. \tag{15}$$

## III. Asymptotic Analysis

We exploit the statistical distribution for the channels $\{\mathbf{H}_{jl}\}$ and the large dimensions of $N$ and $K$ to compute a deterministic approximation of $\gamma_{jk}$ for MRT and RZF, which will be eventually used to find an approximation of the ergodic sum rate. In doing so, we assume the following grow rate of system dimensions:

**Assumption 1.** *The dimensions $N$ and $K$ grow to infinity at the same pace, that is:*

$$0 \leq \liminf K/N \leq \limsup K/N < 1. \quad (16)$$

For technical reasons, the following reasonable assumption is also imposed on the system settings [6]–[8].

**Assumption 2.** *As $N, K \to \infty$*

$$\sup_j \frac{1}{\sqrt{N}} \left\|\overline{\mathbf{H}}_{jj}\right\| < \infty \quad (17)$$

*which implies that the Euclidean norm of the column vectors $\{\overline{\mathbf{h}}_{jjk}\}$ are uniformly bounded in $N, K$.*

Our first result is the asymptotic approximation of the SINR of the system when MRT is employed.

**Theorem 1** (MRT)**.** *Let Assumptions 1 – 2 hold true. If MRT is employed, then $\gamma_{jk}^{\rm MRT} - \overline{\gamma}_{jk}^{\rm MRT} \to 0$ almost surely with*

$$\overline{\gamma}_{jk}^{\rm MRT} = \frac{\overline{\theta}_j \left(\phi_{jjk} + \frac{1}{N}\overline{\mathbf{h}}_{jjk}^H\overline{\mathbf{h}}_{jjk}\right)^2}{\frac{1}{N\rho^{\rm dl}} + \overline{s}_{jk} + \sum\limits_{l=1, l\neq j}^{L} \overline{\theta}_l \phi_{ljk}^2} \quad (18)$$

*where $\overline{s}_{jk}$ takes the form*

$$\overline{s}_{jk} = \frac{1}{N}\sum_{l=1}^{L}\sum_{i=1}^{K}\overline{\theta}_l d_{ljk}\left(\phi_{lli} + \frac{1}{N}\overline{\mathbf{h}}_{lli}^H\overline{\mathbf{h}}_{lli}\right) + \frac{1}{N}\sum_{i=1, i\neq k}^{K}\overline{\theta}_j\left(\phi_{jji}\frac{1}{N}\overline{\mathbf{h}}_{jjk}^H\overline{\mathbf{h}}_{jjk} + \frac{1}{N}\left|\overline{\mathbf{h}}_{jji}^H\overline{\mathbf{h}}_{jjk}\right|^2\right) \quad (19)$$

*and $\overline{\theta}_j$ is given by*

$$\overline{\theta}_j = \left(\frac{1}{K}\sum_{k=1}^{K}\left(\phi_{jjk} + \frac{1}{N}\overline{\mathbf{h}}_{jjk}^H\overline{\mathbf{h}}_{jjk}\right)\right)^{-1}. \quad (20)$$

*Proof:* The proof is omitted for space limitation. It follows the same arguments of those in [3, Theorem 4] taking into account the different system model due to the presence of the LOS components. ∎

As for the RZF, we call $\mathbf{\Phi}_{jj} = {\rm diag}\{\phi_{jj1}, \ldots, \phi_{jjK}\}$ and rewrite $\widehat{\mathbf{H}}_{jj} = [\widehat{\mathbf{h}}_{jj1} \cdots \widehat{\mathbf{h}}_{jjK}] \in \mathbb{C}^{N\times K}$ as

$$\widehat{\mathbf{H}}_{jj} = \mathbf{Z}_{jj}\mathbf{\Phi}_{jj} + \overline{\mathbf{H}}_{jj}. \quad (21)$$

Then, let us introduce the fundamental equations that are needed to express a deterministic equivalent of $\gamma_{jk}$ under RZF. We start with the following set of equations:

$$\delta_j = \frac{1}{N}{\rm tr}\left(\lambda_j\left(1+\widetilde{\delta}_j\right)\mathbf{I}_N + \frac{1}{N}\overline{\mathbf{H}}_{jj}(\mathbf{I}_K + \delta_j\mathbf{\Phi}_{jj})^{-1}\overline{\mathbf{H}}_{jj}^H\right)^{-1} \quad (22)$$

$$\widetilde{\delta}_j = \frac{1}{N}{\rm tr}\mathbf{\Phi}_{jj}\left(\lambda_j\left(\mathbf{I}_K + \delta_j\mathbf{\Phi}_{jj}\right) + \frac{1}{N}\frac{\overline{\mathbf{H}}_{jj}^H\overline{\mathbf{H}}_{jj}}{1+\widetilde{\delta}_j}\right)^{-1} \quad (23)$$

which admits a unique positive solution in the class of Stieltjes transforms of non-negative measures with support $\mathbb{R}_+$ [6], [7]. The matrices

$$\mathbf{T}_j = \left(\lambda_j\left(1+\widetilde{\delta}_j\right)\mathbf{I}_N + \frac{1}{N}\overline{\mathbf{H}}_{jj}(\mathbf{I}_K + \delta_j\mathbf{\Phi}_{jj})^{-1}\overline{\mathbf{H}}_{jj}^H\right)^{-1} \quad (24)$$

$$\widetilde{\mathbf{T}}_j = \left(\lambda_j\left(\mathbf{I}_K + \delta_j\mathbf{\Phi}_{jj}\right) + \frac{1}{N}\frac{\overline{\mathbf{H}}_{jj}^H\overline{\mathbf{H}}_{jj}}{1+\widetilde{\delta}_j}\right)^{-1} \quad (25)$$

are approximations of the resolvent $\mathbf{Q}_{jj} = (\frac{1}{N}\widehat{\mathbf{H}}_{jj}\widehat{\mathbf{H}}_{jj}^H + \lambda_j\mathbf{I}_N)^{-1}$ and co-resolvent $\widetilde{\mathbf{Q}}_{jj} = (\frac{1}{N}\widehat{\mathbf{H}}_{jj}^H\widehat{\mathbf{H}}_{jj} + \lambda_j\mathbf{I}_K)^{-1}$. Also, let us define the following quantities:

$$\vartheta_j = \frac{1}{N}{\rm tr}\left(\mathbf{T}_j^2\right) \quad (26)$$

$$\widetilde{\vartheta}_j = \frac{1}{N}{\rm tr}\left(\mathbf{\Phi}_{jj}\widetilde{\mathbf{T}}_j\right)^2 \quad (27)$$

$$F_j = \frac{1}{N^2}{\rm tr}\left(\mathbf{T}_j^2\overline{\mathbf{H}}_{jj}\left(\mathbf{I}_K + \delta_j\mathbf{\Phi}_{jj}\right)^{-2}\mathbf{\Phi}_{jj}\overline{\mathbf{H}}_{jj}^H\right) \quad (28)$$

$$\Delta_j = (1-F_j)^2 - \lambda_j^2\vartheta_j\widetilde{\vartheta}_j. \quad (29)$$

Then, the following theorem summarizes a major result of this work:

**Theorem 2** (RZF)**.** *Let Assumptions 1 – 2 hold true. Then, if RZF is employed we have that $\max_k\left|\gamma_{jk}^{\rm RZF} - \overline{\gamma}_{jk}^{\rm RZF}\right| \to 0$ almost surely with*

$$\overline{\gamma}_{jk}^{\rm RZF} = \frac{\overline{\psi}_j\left(\frac{\overline{u}_{jk}}{1+\overline{u}_{jk}}\right)^2}{\frac{1}{N\rho^{\rm dl}} + \overline{s}_{jk} + \sum\limits_{l=1, l\neq j}^{L}\overline{\psi}_l\left(\frac{\phi_{ljk}\delta_l}{1+\overline{u}_{lk}}\right)^2} \quad (36)$$

*with $\overline{u}_{jk}$ given by*

$$\overline{u}_{jk} = \frac{1}{\lambda_j[\widetilde{\mathbf{T}}_j]_{kk}} - 1 \quad (37)$$

*whereas $\overline{\psi}_j$ and $\overline{s}_{jk}$ take the form:*

$$\overline{\psi}_j = \left(\frac{\lambda_j^2\vartheta_j}{\Delta_j}\frac{1}{K}{\rm tr}\mathbf{\Phi}_{jj}\widetilde{\mathbf{T}}_j^2 + \frac{1-F_j}{\Delta_j(1+\tilde{\delta}_j)^2}\frac{1}{KN}{\rm tr}\widetilde{\mathbf{T}}_j\overline{\mathbf{H}}_{jj}^H\overline{\mathbf{H}}_{jj}\widetilde{\mathbf{T}}_j\right)^{-1} \quad (38)$$

*and*

$$\overline{s}_{jk} = \sum_{l=1}^{L}\overline{\psi}_l\left(\xi_{ljk} - \lambda_l\mu_{ljk}\right) - \overline{\psi}_j\left(\frac{\overline{u}_{jk}}{1+\overline{u}_{jk}}\right)^2 - \sum_{l=1, l\neq j}^{L}\overline{\psi}_l\left(\frac{\phi_{ljk}\delta_l}{1+\overline{u}_{lk}}\right)^2 \quad (39)$$

*with $\xi_{ljk}$ and $\mu_{ljk}$ being given by* (30) – (35) *on the top of the next page.*

*Proof:* The proof is very much involved and relies on results in random matrix theory [6], [19] as well as some recent ones on deterministic equivalents of bilinear forms [7]. Due to the space limitations, it is omitted. A complete proof will be provided in the extended version. ∎

$$\xi_{ljk} = d_{ljk}\delta_l - \lambda_l[\widetilde{\mathbf{T}}_l]_{kk}\left(\phi_{ljk}\delta_l\right)^2 \qquad l \neq j \tag{30}$$

$$\xi_{jjk} = \delta_j\left(d_{jjk} - \phi_{jjk}\right) + 1 - \lambda_j[\widetilde{\mathbf{T}}_j]_{kk} \qquad l = j \tag{31}$$

---

$$\mu_{ljk} = d_{ljk}\overline{\nu}_l - \phi_{ljk}^2\delta_l\lambda_l[\widetilde{\mathbf{T}}_l]_{kk}\Big(2\overline{\nu}_l - \delta_l\lambda_l[\widetilde{\mathbf{T}}_l]_{kk}(\phi_{llk}\overline{\nu}_l + \overline{\varsigma}_{lk})\Big) \qquad l \neq j \tag{32}$$

$$\mu_{jjk} = \overline{\nu}_j\left(d_{jjk} - \phi_{jjk}\left(1 - \lambda_j^2[\widetilde{\mathbf{T}}_j]_{kk}^2\right)\right) + \lambda_j^2[\widetilde{\mathbf{T}}_j]_{kk}^2\overline{\varsigma}_{jk} \qquad l = j \tag{33}$$

---

$$\overline{\nu}_l = \frac{1}{\Delta_l}\frac{1}{N}\mathrm{tr}\mathbf{T}_l^2 \tag{34}$$

$$\overline{\varsigma}_{lk} = \frac{1-F_l}{\Delta_l}\frac{\left[\widetilde{\mathbf{T}}_l\frac{1}{N}\overline{\mathbf{H}}_{ll}^H\overline{\mathbf{H}}_{ll}\widetilde{\mathbf{T}}_l\right]_{kk}}{\lambda_l^2[\widetilde{\mathbf{T}}_l]_{kk}^2(1+\tilde{\delta}_l)^2} + \frac{\vartheta_l}{\Delta_l}\left(\frac{[\widetilde{\mathbf{T}}_l\mathbf{\Phi}_{ll}\widetilde{\mathbf{T}}_l]_{kk}}{[\widetilde{\mathbf{T}}_l]_{kk}^2} - \phi_{llk}\right) \tag{35}$$

---

### *A. Limiting case $N \to \infty$ with $K/N \to 0$*

We now look at the limiting case in which $N \to \infty$ such that $K/N \to 0$. The following results are easily obtained from the asymptotic analysis above:

**Corollary 1** (MRT). *If $N \to \infty$ such that $K/N \to 0$, then:*

$$\overline{\gamma}_{jk}^{\mathrm{MRT}} = \frac{\overline{\theta}_j\left(\phi_{jjk} + \frac{1}{N}\overline{\mathbf{h}}_{jjk}^H\overline{\mathbf{h}}_{jjk}\right)^2}{\sum\limits_{i=1,i\neq k}^{K}\left|\frac{1}{N}\overline{\mathbf{h}}_{jji}^H\overline{\mathbf{h}}_{jjk}\right|^2 + \sum\limits_{l=1,l\neq j}^{L}\overline{\theta}_l\phi_{ljk}^2} \tag{40}$$

*with $\overline{\theta}_j$ given by* (20).

**Corollary 2** (RZF). *If $N \to \infty$ such that $K/N \to 0$, we have that:*

$$\overline{\gamma}_{jk}^{\mathrm{RZF}} = \frac{\overline{\psi}_j\left(1 - \lambda_j[\widetilde{\mathbf{T}}_j]_{kk}\right)^2}{\overline{s}_{jk} + \sum\limits_{l=1,l\neq j}^{L}\overline{\psi}_l\phi_{ljk}^2[\widetilde{\mathbf{T}}_l]_{kk}^2} \tag{46}$$

*where*

$$\overline{\psi}_j = \frac{1}{K}\mathrm{tr}\mathbf{\Phi}_{jj}\widetilde{\mathbf{T}}_j^2 + \frac{1}{KN}\mathrm{tr}\widetilde{\mathbf{T}}_j\overline{\mathbf{H}}_{jj}^H\overline{\mathbf{H}}_{jj}\widetilde{\mathbf{T}}_j \tag{47}$$

*with*

$$\widetilde{\mathbf{T}}_j = \left(\lambda_j\mathbf{I}_K + \mathbf{\Phi}_{jj} + \frac{1}{N}\overline{\mathbf{H}}_{jj}^H\overline{\mathbf{H}}_{jj}\right)^{-1} \tag{48}$$

*and also*

$$\overline{s}_{jk} = \sum_{l=1}^{L}\overline{\psi}_l\left(\xi_{ljk} - \lambda_l\mu_{ljk}\right) - \overline{\psi}_j\left(1 - \lambda_j[\widetilde{\mathbf{T}}_j]_{kk}\right)^2 - \sum_{l=1,l\neq j}^{L}\overline{\psi}_l\phi_{ljk}^2[\widetilde{\mathbf{T}}_l]_{kk}^2 \tag{49}$$

*with $\xi_{ljk}$ and $\mu_{ljk}$ being given by* (41) – (45) *on the top of the next page.*

*Proof:* If $N \to \infty$ with $K/N \to 0$, then we have that $\delta_j = \lambda_j^{-1}$, $\widetilde{\delta}_j = 0$. Moreover, we have that $\vartheta_j = \lambda_j^{-2}$, $F_j = \widetilde{\vartheta}_j = 0$, $\Delta_j = 1$, $\overline{\nu}_j = \lambda_j^{-2}$. ■

The above corollaries show that when $N$ grows at a faster rate than $K$, differently from Rayleigh fading, the asymptotic interference does not necessarily vanish as it depends on the LOS components. If MRT is considered, it is easily seen from (40) that the remaining interference is due to the intracell interference, which depends on the inner products between different channel vectors $\overline{\mathbf{h}}_{jji}$ and $\overline{\mathbf{h}}_{jjk}$. Similar observations can be made under RZF.

Consider now a system in which $\frac{1}{N}\overline{\mathbf{h}}_{jji}^H\overline{\mathbf{h}}_{jjk} \to 0\ \forall i \neq k$ as $N \to \infty$. This amounts to assuming that the UEs are selected such that the favorable propagation conditions are asymptotically satisfied [20]. Then, we have that:

**Corollary 3** (MRT). *If $N \to \infty$ with $K/N \to 0$ and $\frac{1}{N}\overline{\mathbf{h}}_{jji}^H\overline{\mathbf{h}}_{jjk} \to 0\ \forall i \neq k$, then:*

$$\overline{\gamma}_{jk}^{\mathrm{MRT}} = \frac{\overline{\theta}_j\left(\phi_{jjk} + \frac{1}{N}\overline{\mathbf{h}}_{jjk}^H\overline{\mathbf{h}}_{jjk}\right)^2}{\sum\limits_{l=1,l\neq j}^{L}\overline{\theta}_l\phi_{ljk}^2} \tag{50}$$

*with $\overline{\theta}_j$ given by* (20).

**Corollary 4** (RZF). *If $N \to \infty$ with $K/N \to 0$ and $\frac{1}{N}\overline{\mathbf{h}}_{jji}^H\overline{\mathbf{h}}_{jjk} \to 0\ \forall i \neq k$, then:*

$$\overline{\gamma}_{jk}^{\mathrm{RZF}} = \frac{\overline{\psi}_j\left(\frac{\phi_{jjk}+\frac{1}{N}\overline{\mathbf{h}}_{jjk}^H\overline{\mathbf{h}}_{jjk}}{\lambda_j+\phi_{jjk}+\frac{1}{N}\overline{\mathbf{h}}_{jjk}^H\overline{\mathbf{h}}_{jjk}}\right)^2}{\sum\limits_{l=1,l\neq j}^{L}\overline{\psi}_l\left(\frac{\phi_{ljk}}{\lambda_l+\phi_{llk}+\frac{1}{N}\overline{\mathbf{h}}_{llk}^H\overline{\mathbf{h}}_{llk}}\right)^2} \tag{51}$$

$$= \frac{\overline{\psi}_j\left(\phi_{jjk} + \frac{1}{N}\overline{\mathbf{h}}_{jjk}^H\overline{\mathbf{h}}_{jjk}\right)^2}{\sum\limits_{l=1,l\neq j}^{L}\left(\frac{\lambda_j+\phi_{jjk}+\frac{1}{N}\overline{\mathbf{h}}_{jjk}^H\overline{\mathbf{h}}_{jjk}}{\lambda_l+\phi_{llk}+\frac{1}{N}\overline{\mathbf{h}}_{llk}^H\overline{\mathbf{h}}_{llk}}\right)^2\overline{\psi}_l\phi_{ljk}^2} \tag{52}$$

*with $\psi_j$ given by*

$$\overline{\psi}_j = \frac{1}{K}\sum_{k=1}^{K}\frac{\phi_{jjk} + \frac{1}{N}\overline{\mathbf{h}}_{jjk}^H\overline{\mathbf{h}}_{jjk}}{\left(\lambda_j + \phi_{jjk} + \frac{1}{N}\overline{\mathbf{h}}_{jjk}^H\overline{\mathbf{h}}_{jjk}\right)^2}. \tag{53}$$

$$\xi_{ljk} = \frac{d_{ljk}}{\lambda_l} - \frac{[\widetilde{\mathbf{T}}_l]_{kk}}{\lambda_l}\phi_{ljk}^2 \qquad l \neq j \tag{41}$$

$$\xi_{jjk} = \frac{1}{\lambda_j}(d_{jjk} - \phi_{jjk}) + 1 - \lambda_j[\widetilde{\mathbf{T}}_j]_{kk} \qquad l = j \tag{42}$$

---

$$\mu_{ljk} = \frac{1}{\lambda_l^2}\Big(d_{ljk} - 2\phi_{ljk}^2[\widetilde{\mathbf{T}}_l]_{kk}\Big) + \phi_{ljk}^2\Big(\frac{\phi_{llk}}{\lambda_l^2} + \overline{\varsigma}_{lk}\Big)[\widetilde{\mathbf{T}}_l]_{kk}^2 \qquad l \neq j \tag{43}$$

$$\mu_{jjk} = \frac{1}{\lambda_j^2}\Big(d_{jjk} - \phi_{jjk}(1 - \lambda_j^2[\widetilde{\mathbf{T}}_j]_{kk}^2)\Big) + \lambda_j^2[\widetilde{\mathbf{T}}_j]_{kk}^2\overline{\varsigma}_{jk} \qquad l = j \tag{44}$$

---

$$\overline{\varsigma}_{lk} = \frac{\Big[\widetilde{\mathbf{T}}_l\frac{1}{N}\overline{\mathbf{H}}_{ll}^H\overline{\mathbf{H}}_{ll}\widetilde{\mathbf{T}}_l\Big]_{kk}}{\lambda_l^2[\widetilde{\mathbf{T}}_l]_{kk}^2} + \frac{1}{\lambda_l^2}\left(\frac{[\widetilde{\mathbf{T}}_l\mathbf{\Phi}_{ll}\widetilde{\mathbf{T}}_l]_{kk}}{[\widetilde{\mathbf{T}}_l]_{kk}^2} - \phi_{llk}\right) \tag{45}$$

---

*Proof:* The result follows from observing that, if $\frac{1}{N}\overline{\mathbf{h}}_{jji}^H\overline{\mathbf{h}}_{jjk} \to 0 \;\forall i \neq k$, then $\widetilde{\mathbf{T}}_j$ in (48) becomes diagonal with elements $[\widetilde{\mathbf{T}}_j]_{kk} = (\lambda_j + \phi_{jjk} + \frac{1}{N}\overline{\mathbf{h}}_{jjk}^H\overline{\mathbf{h}}_{jjk})^{-1}$. Also, it can be proved after standard but tedious calculus that $\overline{s}_{jk} \to 0$. ∎

In line with [15], [20], the above corollaries show that if the MIMO Rician fading channel results in favorable propagation, then the interference vanishes as $N$ grows unbounded for both MRT and RZF. In practice, this means that those asymptotic values can only be achieved if some UEs are dropped from service [21]. From Corollaries 3 and 4, it also turns out that, as for Rayleigh fading channels [3, Corollary 1], the asymptotic SINRs under RZF and MRT are not necessarily the same. This is because the matrix $\mathbf{Q}_j$ under RZF depends on the correlation matrix $\mathbf{\Phi}_{jj}$ through (21).

### *B. A case study*

The above asymptotic results will be validated in Section IV and used to make comparisons between MRT and RZF under different settings. In addition to this, they can be used to get some insights on the effects of the Rician component. To this end, let us consider the regime in which $N$ and $K$ grow to infinity at the same pace and assume that the channel can be simply modelled as follows:

$$\mathbf{h}_{jjk} = \sqrt{\frac{1}{1+\kappa}}\mathbf{z}_{jjk} + \sqrt{\frac{\kappa}{1+\kappa}}\mathbf{a}_{jjk} \qquad l = j \tag{54}$$

$$\mathbf{h}_{jlk} = \sqrt{\alpha}\mathbf{z}_{jlk} \qquad l \neq j \tag{55}$$

such that

$$\phi_{jjk} = \frac{1}{(1+\kappa)^2}\nu \qquad l = j \tag{56}$$

$$\phi_{jlk} = \frac{\alpha}{1+\kappa}\nu \qquad l \neq j \tag{57}$$

with $\nu = \frac{\rho^{\rm tr}}{1+\rho^{\rm tr}\overline{L}}$ and $\overline{L} = \alpha(L-1) + \frac{1}{1+\kappa}$. Assume also that $\mathbf{H}_{jj}$ has orthogonal columns such that $\overline{\mathbf{H}}_{jj}^H\overline{\mathbf{H}}_{jj} = \frac{\kappa}{\kappa+1}\mathbf{I}_K$. This is achieved if the vectors $\mathbf{a}_{jjk}$ are such that $\frac{1}{N}\mathbf{a}_{jjk}^H\mathbf{a}_{jjk} = 1$ and $\frac{1}{N}\mathbf{a}_{jji}^H\mathbf{a}_{jjk} = 0 \;\forall i \neq k$. For simplicity, we consider only MRT (despite being much more involved, similar results can be obtained for RZF):

**Corollary 5** (MRT)**.** *Let Assumptions 1 – 2 hold true. If the channel is modeled as in* (54) *and* $\overline{\mathbf{H}}_{jj}$ *satisfies the condition above* $\forall j$*, then*

$$\overline{\gamma}_{jk}^{\rm MRT} = \frac{1}{\frac{1}{\nu N\rho^{\rm dl}}\frac{1+\kappa}{\tau} + \frac{K}{N\nu}\Big(\overline{L}\frac{1+\kappa}{\tau} + \frac{1}{\tau^2}\frac{\kappa}{1+\kappa}\Big) + \frac{\alpha}{\tau^2}\Big(\overline{L} - \frac{1}{1+\kappa}\Big)}$$

$$= \frac{1}{\underbrace{\frac{\overline{L}}{N\rho^{\rm dl}}\frac{1+\kappa}{\tau}}_{\text{Noise}} + \underbrace{\frac{1}{\rho^{\rm tr}}A}_{\text{Imperfect CSI}} + \underbrace{\frac{K}{N}\overline{L}B}_{\text{Interference}} + \underbrace{\frac{\alpha}{\tau^2}\Big(\overline{L} - \frac{1}{1+\kappa}\Big)}_{\text{Pilot Contamination}}}$$

*where* $\nu = \frac{\rho^{\rm tr}}{1+\rho^{\rm tr}\overline{L}}$ *and* $\overline{L} = \alpha(L-1) + \frac{1}{1+\kappa}$, $\tau = \frac{1}{1+\kappa} + \frac{\kappa}{\nu}$ *and*

$$A = \left(\frac{K}{N}\overline{L} + \frac{1}{N\rho^{\rm dl}}\right)\frac{1+\kappa}{\tau} + \frac{K}{N}\frac{1}{\tau^2}\frac{\kappa}{1+\kappa}$$

$$B = \overline{L}\frac{1+\kappa}{\tau} + \frac{1}{\tau^2}\frac{\kappa}{1+\kappa}.$$

From the above results, it follows that the effective SNR $\nu N\rho^{\rm dl}\frac{\tau}{1+\kappa}$ increases linearly with $N$ as it happens for Rayleigh fading channels [3, Corollary 2]. Also, it increases with the Rician component $\kappa$ as $\nu\frac{\tau}{1+\kappa}$. As for a Rayleigh model, the channel estimation errors and interference vanish only if $N$ grows large. Indeed, if $\kappa$ increases $A$ tends to $\frac{K}{N}\overline{L} + \frac{1}{N\rho^{\rm dl}}$ whereas $B$ goes to $\overline{L}$. On the other hand, the pilot contamination term goes to zero with $\kappa$ as $1/\kappa^2$ (since $\tau \to \kappa/\nu$ as $\kappa$ grows large).

## IV. Numerical results

Monte-Carlo simulations are now used to validate the accuracy of the above asymptotic analysis for finite values of $N$ and $K$. We consider a multi-cell system with $L = 3$ cells. The inner cell radius is normalized to one and we assume that the large scale coefficients are given by $\beta_{jlk} = \frac{1}{x_{jlk}^{\alpha}}$ where $x_{jlk}$ is the distance from BS $j$ to UE $k$ in cell $l$ and $\alpha = 2.5$ is the path loss exponent. We further assume that $\rho^{\rm tr} = 6$ dB and the transmit SNR is $\rho^{\rm dl} = 10$ dB.

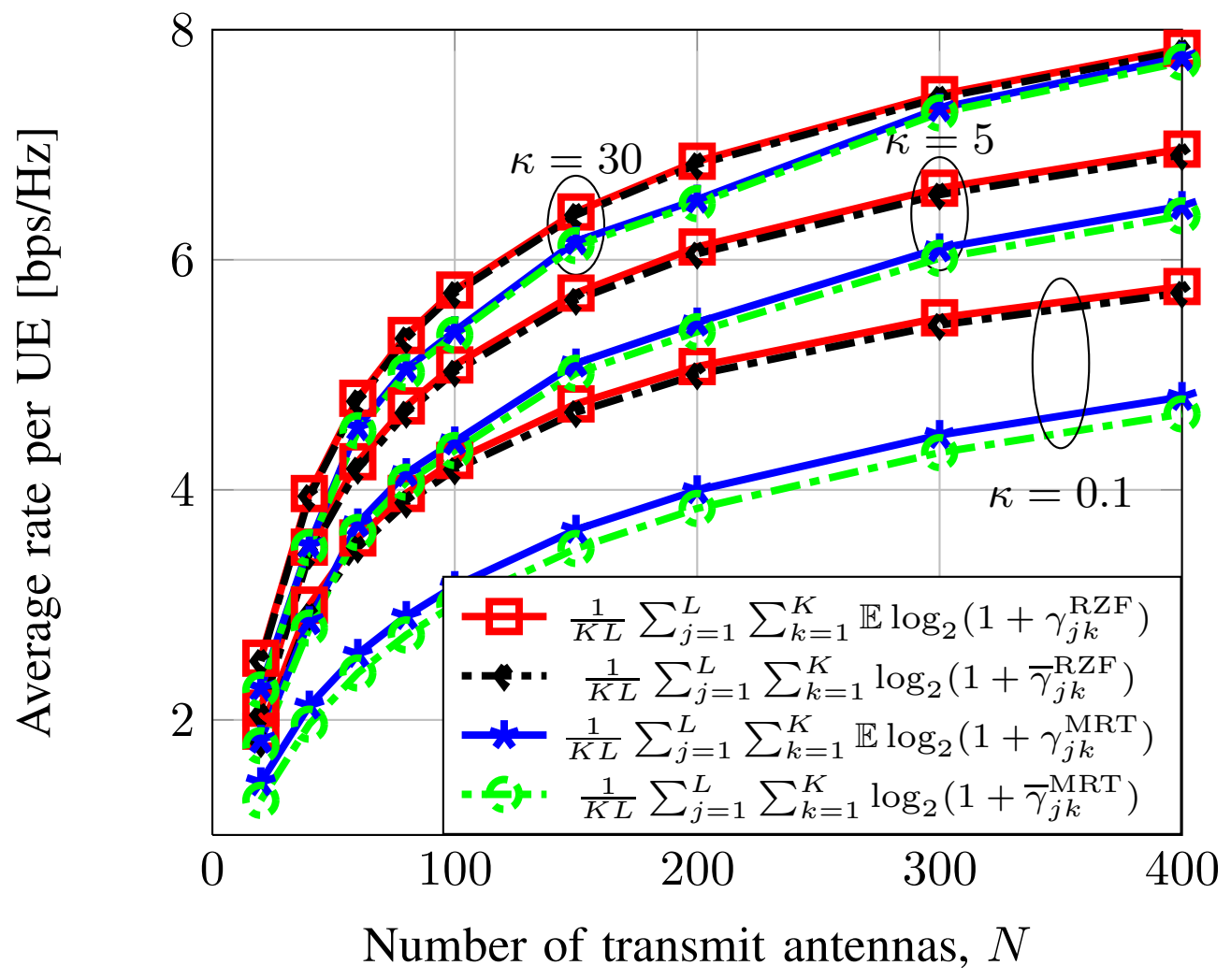

Fig. 1. Average rate per UE vs $N$ when $K = 10$ and the Rician factor is 0.1, 5 and 30.

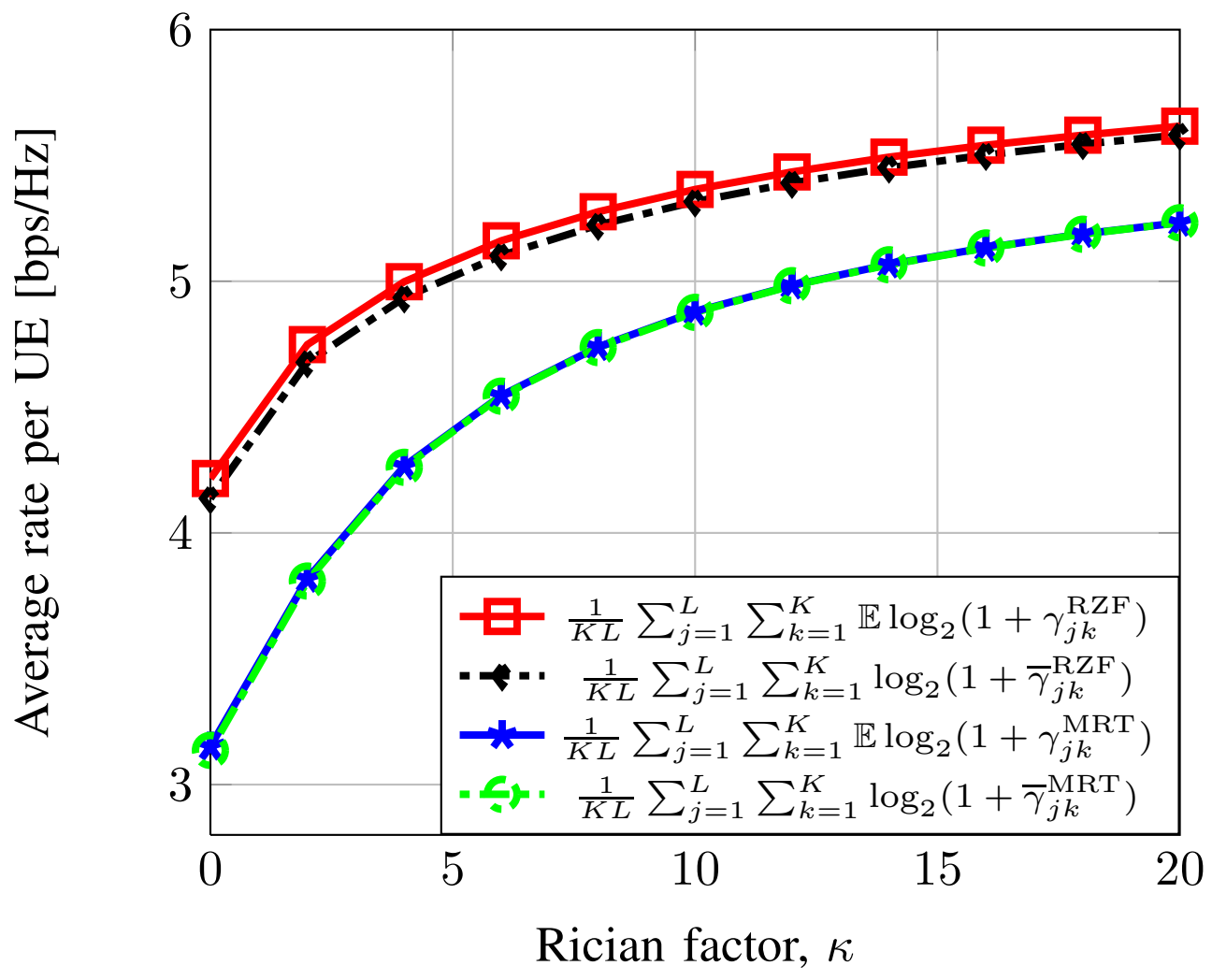

Fig. 2. Average rate per UE vs $\kappa$ when $K = 10$ and $N = 100$.

Fig. 1 illustrates the average rate per UE when $N$ grows large and $K$ is kept fixed to 10. Also, we assume that $\lambda_j^{\rm dl} = \frac{K}{N\rho^{\rm dl}}$. The Rician factor is assumed to be the same for all UEs, i.e., $\kappa_{jk} = \kappa \ \forall j,k$, and equal to $\kappa = 0.1, 5$ and 30. As seen, the asymptotic results perfectly match the Monte Carlo simulations. Moreover, as both $\kappa$ and $N$ increase, the gain of RZF over MRT becomes less significant. In particular, for $\kappa = 30$, RZF and MRT achieve the same performance when $N$ is larger than 300.

In Fig. 2, we further investigate the impact of $\kappa$ when $N$ is kept fixed to 100 and $K = 10$. As expected, both schemes have better better performance for higher values of $\kappa$ and RZF always outperforms MRT.

## V. Conclusions

We investigated the effect of a Rician fading channel on the DL ergodic sum rate of MRT and ZF precoding schemes in massive MIMO systems with a very general channel model under the assumption that data transmission was affected by channel estimation errors and pilot contamination. Recent results from random matrix theory were used to find asymptotic approximations for the sum rate of the investigated precoding schemes. Such approximations turned out to depend only on the long-term channel statistics, the Rician factor and the deterministic component. Numerical results indicated that these approximations are very accurate. Applied to practical networks, such results may lead to important insights into the system behavior, especially with respect to the Rician factors, the deterministic components, CSI quality and induced interference. Moreover, they can be used to simulate the network behavior without to carry out extensive Monte-Carlo simulations.